\documentclass[final,aps,prb,twocolumn,superscriptaddress,showpacs]{revtex4-1}

\usepackage{graphicx}
\usepackage{dcolumn}
\usepackage{amsmath,bm}
\usepackage{color}%
\usepackage{multirow}
\usepackage{array} 
\usepackage{esvect}
\usepackage{braket}
\usepackage{makecell}
\usepackage[utf8]{inputenc}
\usepackage{kotex}
\usepackage{xr}
\usepackage{soul,xcolor}
\setstcolor{red} 

\makeatletter
\newcommand*{\addFileDependency}[1]{
  \typeout{(#1)}
  \@addtofilelist{#1}
  \IfFileExists{#1}{}{\typeout{No file #1.}}
}
\makeatother

\newcommand*{\myexternaldocument}[1]{%
    \externaldocument{#1}%
    \addFileDependency{#1.tex}%
    \addFileDependency{#1.aux}%
}

\myexternaldocument{SI}  

\begin{document}

\title{Temperature--Composition Mapping and Growth-History Dependence of rBN-Related Raman Responses in BN Grown on Fe--B under Flowing N$_2$}

\author{Donghoi Kim}
\affiliation{Department of Information Display, Kyung Hee University, Seoul 02447, Republic of Korea}

\author{Ayoung Cho}
\author{Kyeongho Park}
\affiliation{Department of Physics and Research Institute for Basic Sciences, Kyung Hee University, Seoul 02447, Republic of Korea}

\author{Young Duck Kim}
\affiliation{Department of Physics and Research Institute for Basic Sciences, Kyung Hee University, Seoul 02447, Republic of Korea}

\author{Chinkyo Kim}
\email[Corresponding author. E-mail: ]{ckim@khu.ac.kr}
\affiliation{Department of Information Display, Kyung Hee University, Seoul 02447, Republic of Korea}
\affiliation{Department of Physics and Research Institute for Basic Sciences, Kyung Hee University, Seoul 02447, Republic of Korea}

\date{\today}
\begin{abstract}
BN products formed on Fe--B sources under flowing N$_2$ were investigated primarily at 1250--1400~$^\circ$C over nominal B contents of 1--10~wt\% to determine how rBN-related Raman character depends on growth condition and history. Spatially resolved Raman mapping across 44 growth conditions identified a restricted temperature--composition region in which the integrated rBN-related low-frequency response over 750--835~cm$^{-1}$, $A_{750-835}$, is strongly developed together with the integrated total response of the main $\sim$1367~cm$^{-1}$ BN band, $A_{1367}^{\mathrm{total}}$. In addition, the asymmetric high-wavenumber extension of the $\sim$1367~cm$^{-1}$ band was quantified by subtracting a symmetric fitted core and integrating the far high-wavenumber residual, $T_{\mathrm{high}}^{\mathrm{far}}$. The independently defined $A_{750-835}$ and $T_{\mathrm{high}}^{\mathrm{far}}$ responses exhibit a strong correspondence across the full dataset, supporting their use as complementary rBN-related Raman observables. In contrast, $A_{750-835}/A_{1367}^{\mathrm{total}}$ varies systematically among growth conditions, indicating that the relative prominence of the low-frequency response changes with temperature and composition rather than tracking the overall BN response through a fixed proportionality. Growth-duration experiments at 1350~$^\circ$C further show that this ratio also depends on high-temperature N$_2$ dwell time, establishing growth history as an additional kinetic variable. Powder X-ray diffraction on selected products independently confirms the presence of rBN-related stacking, while second-harmonic generation provides complementary symmetry-sensitive evidence. These results define a processing window favorable for strong rBN-related Raman response and introduce a quantitative framework for tracking both the low-frequency and band-asymmetry signatures associated with rBN-related BN structure.
\end{abstract}

\maketitle

\section{Introduction}

Rhombohedral and hexagonal boron nitride consist of the same sp$^2$-bonded BN sheets but differ in their interlayer registry. The ABC stacking of rBN produces crystallographic symmetry and stacking-sensitive physical properties distinct from those of AA$'$-stacked hBN, including different diffraction, vibrational, and optical responses.\cite{Moret-APL-119-262102,Wang-Nature-629-74,Qi-AM-36-2303122} Accordingly, controlling layered-BN synthesis requires not only forming crystalline BN sheets but also selecting and preserving the stacking sequence during growth.

Several routes have been developed to access rBN. Ammonothermal synthesis has produced rBN-containing products at 750~$^\circ$C under high-pressure NH$_3$,\cite{Maruyama-MaterLett-232-110} and more recent work reported predominantly rBN on nitrided NiCr alloy surfaces.\cite{Dooley-PSSB-262-2500170} Epitaxial vapor-phase growth has achieved rBN near 1500~$^\circ$C on AlN-buffered sapphire and on ZrB$_2$(0001)/SiC,\cite{Chubarov-CGD-12-3215,Chubarov-ChemMater-27-1640,Souqui-JVSTA-39-013405} while vicinal FeNi(111) and Ni surfaces have enabled stacking control through engineered step structures.\cite{Qi-AM-36-2303122,Wang-Nature-629-74} These approaches demonstrate effective control of rhombohedral stacking but generally rely on high pressure, high temperature, epitaxial templates, or deliberately structured single-crystal metal surfaces.

Fe-based growth systems provide a complementary route in which BN stacking varies strongly with composition and thermal history. Desrat \textit{et al.} obtained predominantly rhombohedral bulk crystallites from a B- and N-saturated Fe--2.5~wt\% B flux at 1550--1600~$^\circ$C followed by extremely slow cooling,\cite{Desrat-PhysRevMater-9-124001} whereas Shi \textit{et al.} reported multilayer hBN on solid Fe$_2$B under flowing N$_2$ at 1100--1300~$^\circ$C and from molten Fe$_{82}$B$_{18}$ at 1250~$^\circ$C through isothermal growth processes.\cite{Shi-RSCAdv-9-10155,Shi-NC-11-849} These contrasting results indicate that BN stacking in Fe--B systems depends on the combined effects of temperature, alloy composition, and growth history.

Raman spectroscopy provides a practical means to examine how BN formation and rBN-related stacking character evolve with growth conditions, but their dependence on Fe--B composition, temperature, and growth history has not been systematically established. Here, we investigate BN formation on Fe--B sources under flowing N$_2$ over a broad temperature--composition range using spatially resolved Raman mapping, complemented by growth-duration experiments, powder X-ray diffraction, and polarization-dependent second-harmonic generation. A key contribution of this work is to quantify the previously reported high-wavenumber asymmetry of the main BN Raman band\cite{Desrat-PhysRevMater-9-124001} as a far high-wavenumber residual response and compare it directly with the low-frequency rBN-related response. By combining these complementary Raman observables with the overall BN response, we identify a restricted temperature--composition window in which rBN-related spectral character is strongly developed and show that this response is further modulated by growth history. These results highlight the evolving Fe--B source environment as an important factor influencing the final BN structural response.

\section{Methods}

\subsection{Fe--B source preparation and thermal treatment}

Fe--B sources were prepared by mixing Fe and B powders at selected weight ratios corresponding to nominal B contents of 1--10~wt\%. A total of 4~g of powder mixture was used for each growth experiment. The powder mixtures were loaded into an alumina boat (Al$_2$O$_3$ purity: 99.8\%, dimensions:~20$\times$25$\times$10~mm (L$\times$W$\times$H)) and heated in a horizontal tube furnace under Ar/H$_2$ (300/50~sccm) at 10~$^\circ$C~min$^{-1}$ to the target temperature. After annealing for 60~min, the atmosphere was switched to N$_2$ (300~sccm) and BN growth was carried out for 60~min, followed by cooling at 10~$^\circ$C~min$^{-1}$ under N$_2$. The primary growth-temperature range was 1250--1400~$^\circ$C. For growth-duration experiments, 8, 8.8, and 10~wt\% B sources were treated at 1350~$^\circ$C with N$_2$ growth durations of 60 or 300~min; all other thermal conditions were unchanged.

\subsection{Macroscopic source morphology and phase-diagram reference}

Post-growth sources were classified into five groups (A--E) according to their macroscopic surface appearance and apparent film-like coverage. The overall shape of the Fe--B source was recorded as an ancillary observation but was not used as a classification criterion. White surface regions were classified optically as apparent film-like coverage, while BN-specific structural responses were evaluated independently by Raman spectroscopy and powder X-ray diffraction. Approximate Fe--B phase boundaries from evaluated phase-diagram data were overlaid on the temperature--composition map as references for the nominal alloy conditions.\cite{Liao-PD-41-1993,Okamoto-JPED-25-297,Rompaey-JAC-334-173}

\subsection{Raman spectroscopy and mapping}

Raman measurements were performed using a Renishaw inVia microscope with 532~nm excitation and an 1800~lines~mm$^{-1}$ grating. Spectra were acquired over approximately 716--2366~cm$^{-1}$ with a 1~s acquisition time per pixel and a laser power of 2.64~mW. For each specimen, nine spatially separated locations arranged in an approximately \(3\times3\) grid with $\sim$2~mm spacing were measured. At each location, a \(7\times7\)-pixel Raman map covering approximately \(49~\mu\mathrm{m}^2\) was acquired. The same instrument-software baseline subtraction was applied to all spectra, with no additional baseline or offset correction.

The low-frequency Raman response was quantified as

$$
A_{750-835}
=
\int_{750}^{835} I_{\mathrm{exp}}(\omega)\,d\omega .
$$

The 750--835~cm$^{-1}$ window was selected to encompass the broad 760--825~cm$^{-1}$ feature reported for predominantly rBN crystallites.\cite{Desrat-PhysRevMater-9-124001}

For the BN Raman band near 1367~cm$^{-1}$, the peak maximum was first located within 1355--1375~cm$^{-1}$, and a symmetric pseudo-Voigt function was fitted over the central $\pm10$~cm$^{-1}$ region. Using the fitted peak center, $\omega_0$, the total experimental BN-band response was defined as

$$
A_{1367}^{\mathrm{total}}
=
\int_{\omega_0-100}^{\omega_0+150}
I_{\mathrm{exp}}(\omega)\,d\omega .
$$

To quantify the extended high-wavenumber response of the 1367-cm$^{-1}$ band, the residual relative to the fitted symmetric core was defined as

$$
R(\omega)
=
I_{\mathrm{exp}}(\omega)-I_{\mathrm{core}}(\omega).
$$

The far high-wavenumber residual response was then calculated as

$$
T_{\mathrm{high}}^{\mathrm{far}}
=
\int_{\omega_0+50}^{\omega_0+150}
R(\omega)\,d\omega .
$$

Signed residual values were retained without truncation.

Acquisition artifacts identified by inspection of the spectral overlays were excluded manually, whereas low core-fit $R^2$ values were retained as review flags rather than used as automatic exclusion criteria. No specimen-level exclusion was applied, and all 44 growth conditions were retained in the analysis.

For each \(7\times7\) local map, the median values of \(A_{750-835}\), \(A_{1367}^{\mathrm{total}}\), and \(T_{\mathrm{high}}^{\mathrm{far}}\) were calculated from analysis-valid pixels. For each specimen, the specimen-level Raman quantity was calculated as the arithmetic mean of the available map-level medians from the nine spatially distributed Raman maps. Maps lacking sufficient analysis-valid spectra after exclusion of genuine spectral or acquisition artifacts were not included in the specimen-level average. The standard deviation, minimum, and maximum of the available map-level medians were retained to characterize within-specimen spatial heterogeneity. The nine map locations were treated as spatial sampling positions rather than as independent growth replicates.  For the 1250~$^\circ$C/5~wt\% B specimen, only three maps provided sufficient analysis-valid spectra because unidentified side peaks precluded reliable symmetric-core residual analysis in the remaining maps.

\subsection{Powder X-ray diffraction}

Powder X-ray diffraction (XRD) was performed on mechanically collected BN-containing products using a PANalytical X'Pert PRO MPD diffractometer with Cu K$\alpha$ radiation (45~kV, 30~mA). Powder XRD patterns were acquired over $2\theta=20$--60$^\circ$ with a 0.020$^\circ$ step size and a scan rate of approximately 0.0308$^\circ$~s$^{-1}$.

The hBN(100) and rBN(012) reflections near 41.6$^\circ$ and 45.6$^\circ$, respectively, were fitted with local backgrounds. The relative diffraction descriptor was defined as
\[
Q_{\mathrm{XRD}}
=
\frac{I_{\mathrm{rBN(012)}}}
{I_{\mathrm{hBN(100)}}},
\]
where the intensities denote integrated fitted peak areas. The rBN-related reflection near 42.6$^\circ$ was not used because of overlap with Fe--B-related diffraction.

\subsection{Second-harmonic generation measurements}

A wet-etch-separated BN product grown at 1350~$^\circ$C with 4.3~wt\% B was examined by reflection-mode second-harmonic generation (SHG). An 800~nm femtosecond excitation beam generated using a PHAROS laser coupled to an ORPHEUS optical parametric amplifier was focused onto the sample with a 100$\times$ objective (NA = 0.75). The excitation power was 100~$\mu$W. Polarization was controlled with a half-wave plate, and the reflected SHG signal near 400~nm was polarization analyzed and recorded using an IsoPlane SCT-320 spectrograph equipped with a PyLoN 2KBUV CCD detector, a 1200~lines~mm$^{-1}$ grating, and an exposure time of 5~s.

\section{Results and discussion}

\subsection{Temperature--composition dependence of macroscopic surface characteristics}

\begin{figure*}
\includegraphics[width=2.0\columnwidth]{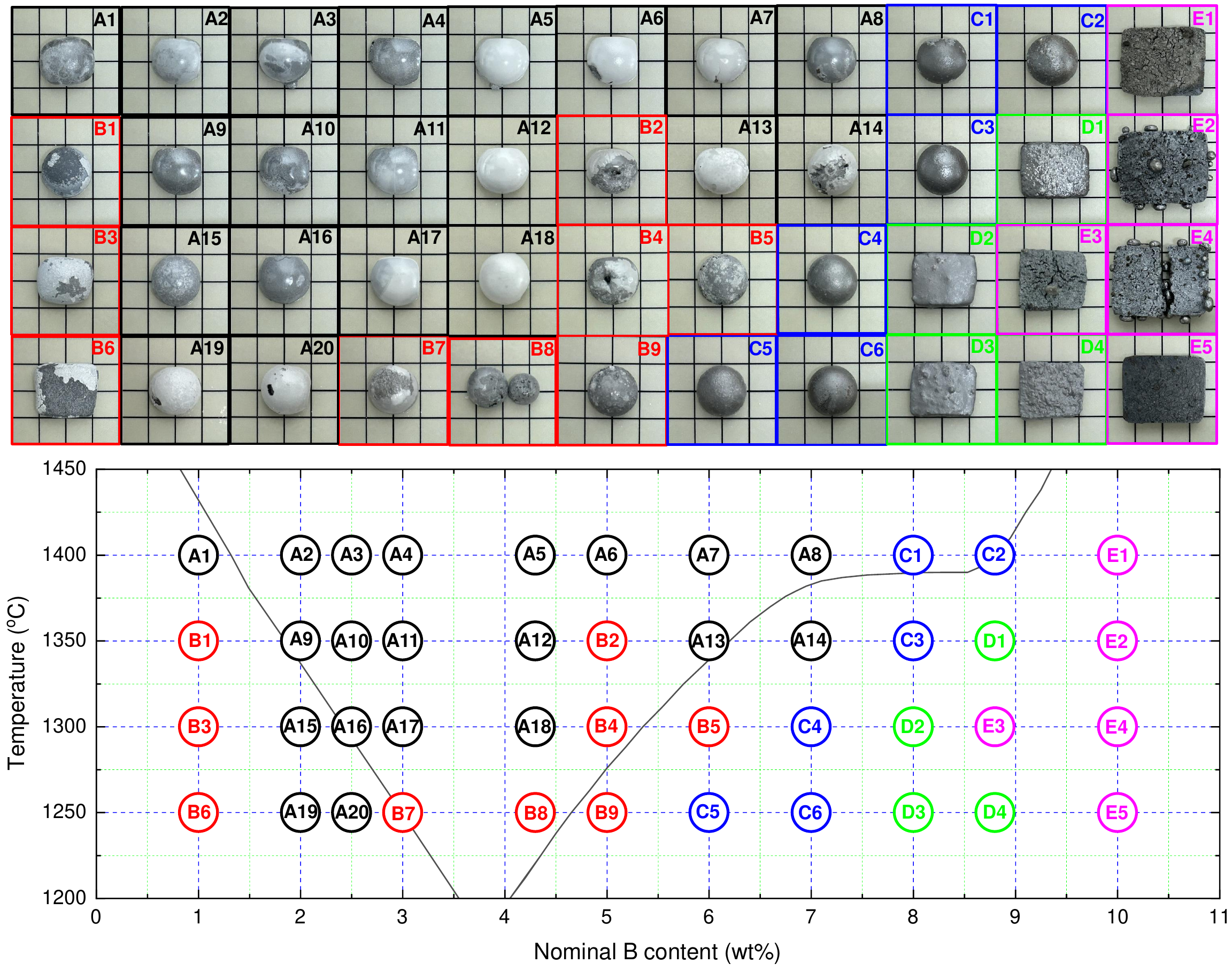}
\caption{
Macroscopic appearance of the Fe--B sources after growth, displayed together with their nominal positions in the temperature--composition space. The lower diagram shows the growth conditions as a function of growth temperature and nominal B content in the Fe--B source, and the optical photographs show the corresponding sources after growth. In each photograph, the background grid provides a size reference, with four grid squares corresponding to 26~mm in both the horizontal and vertical directions. The dashed curves indicate approximate Fe--B phase-boundary guides and provide thermodynamic context for the nominal alloy conditions. The specimens are classified into five groups (A--E) according to their macroscopic surface appearance and apparent film-like coverage: (A) extensive apparent film-like surface coverage; (B) partial apparent film-like surface coverage; (C) metallic surface appearance without apparent film-like coverage; (D) smooth surface appearance without apparent film-like coverage; and (E) rough surface appearance without apparent film-like coverage. The overall shape of the Fe--B source is not used as a classification criterion. White surface regions are classified optically as apparent film-like coverage, whereas the BN-specific structural response is evaluated independently by Raman spectroscopy and powder X-ray diffraction. The phase-boundary overlay is based on evaluated Fe--B phase-diagram information \cite{Liao-PD-41-1993,Okamoto-JPED-25-297,Rompaey-JAC-334-173}.
}
\label{fig:sample-images}
\end{figure*}

The macroscopic appearance of the Fe--B sources after growth was first examined to establish the physical context for the subsequent structural analyses, as summarized in Fig.~\ref{fig:sample-images}. The post-growth sources exhibit systematic variations in surface appearance and apparent film-like coverage across the explored temperature--composition space. Based on these surface characteristics, the specimens were classified into five groups, A--E, as defined in Methods and summarized in the figure caption. The resulting classification provides a compact description of the macroscopic surface products and allows their evolution to be compared with growth temperature, nominal B content, and the approximate Fe--B phase-boundary guides.

The A--E distribution changes systematically across the temperature--composition map, indicating that the macroscopic surface state of the source is strongly dependent on the growth condition. Regions exhibiting extensive optically film-like surface coverage are concentrated within a distinct part of the processing space, whereas progressively different surface appearances occur outside this region. Although the overall source shape also changes across the map, these shape variations are treated here as ancillary observations rather than as a basis for classification. The proximity of changes in macroscopic surface appearance to the approximate Fe--B phase-boundary guides provides useful thermodynamic context, but the optical observations alone do not establish the microscopic phase state of the source.

The optical classification therefore serves as an independent macroscopic measure of how the source surface evolves across the growth space, while the Raman measurements provide BN-specific structural information. As shown below, the A--E classification is closely reflected in the magnitudes of the low-frequency and total BN Raman responses, providing a direct connection between the macroscopic growth-product distribution and the spectroscopic growth map.

\subsection{Raman observables for separating overall BN response from relative structural character}

\begin{figure}
\includegraphics[width=1.0\columnwidth]{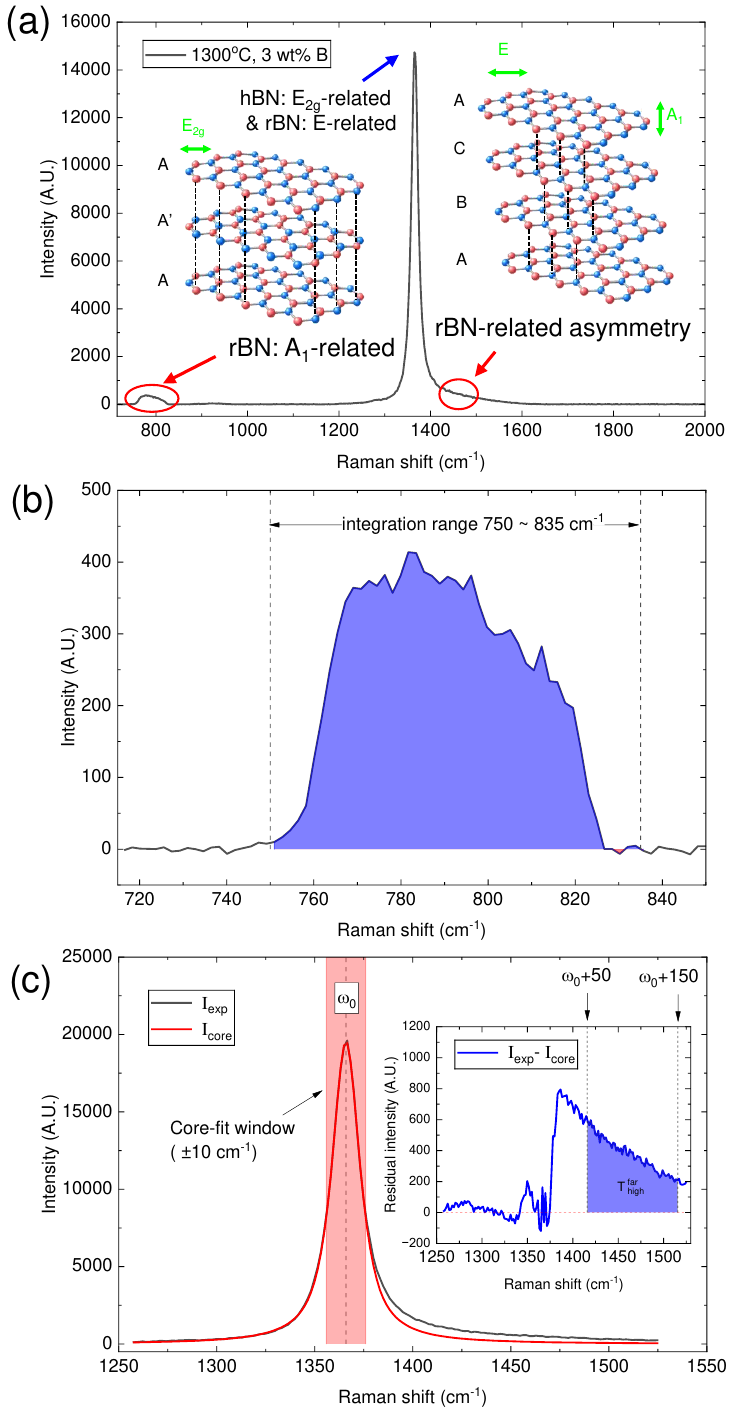}
\caption{
Representative Raman spectrum and definition of the Raman responses used in the analysis. (a) Raman spectrum obtained from the sample grown at 1300~$^\circ$C with 3~wt\% B, shown together with schematic AA$'$-stacked hBN and ABC-stacked rBN structures and the characteristic in-plane and out-of-plane vibrational directions associated with the hBN $E_{2g}$, rBN $E$, and rBN $A_1$ modes. The main BN band near 1367~cm$^{-1}$ is associated with the in-plane hBN $E_{2g}$ and rBN $E$ responses, whereas the low-frequency 750--835~cm$^{-1}$ region contains the rBN $A_1$-related response. (b) Enlarged low-frequency region. The shaded area denotes the integrated low-frequency Raman response, $A_{750-835}$, evaluated from 750 to 835~cm$^{-1}$. (c) Experimental 1367-cm$^{-1}$ band, $I_{\mathrm{exp}}$, and the symmetric pseudo-Voigt core, $I_{\mathrm{core}}$, obtained by fitting the central $\pm10$~cm$^{-1}$ region around the initially located peak maximum. The dashed vertical line indicates the fitted peak center, $\omega_0$. The total experimental BN-band response, $A_{1367}^{\mathrm{total}}$, is obtained by integrating $I_{\mathrm{exp}}$ from $\omega_0-100$ to $\omega_0+150$~cm$^{-1}$. The inset shows the signed residual, $I_{\mathrm{exp}}-I_{\mathrm{core}}$; the shaded region between $\omega_0+50$ and $\omega_0+150$~cm$^{-1}$ defines the far high-wavenumber residual response, $T_{\mathrm{high}}^{\mathrm{far}}$.
}
\label{fig:rBN-related-Raman-feature}
\end{figure}

Raman spectroscopy and spatially resolved Raman mapping were used to examine rBN-related spectral signatures across the temperature--composition space on the unground specimen surfaces. Desrat \textit{et al.} reported two characteristic Raman features in predominantly rhombohedral BN crystallites: a broad band extending approximately from 760 to 825~cm$^{-1}$ and an asymmetric high-wavenumber tail of the main BN mode near 1365--1366~cm$^{-1}$.\cite{Desrat-PhysRevMater-9-124001} The former was interpreted in connection with the out-of-plane $A_1$-related response of rBN, whereas the microscopic origin of the extended high-wavenumber tail remains less firmly established. Guided by these observations, we quantified the low-frequency response, the total BN Raman response, and the far high-wavenumber residual response directly from the measured spectra. Figure~\ref{fig:rBN-related-Raman-feature} illustrates the analysis using a representative spectrum from the specimen grown at 1300~$^\circ$C with 3~wt\% B. The spectrum exhibits the general sp$^2$-BN band near 1367~cm$^{-1}$ together with a distinct low-frequency response over 750--835~cm$^{-1}$ [Fig.~\ref{fig:rBN-related-Raman-feature}(a)].

The broad low-frequency response is quantified by integrating the Raman intensity from 750 to 835~cm$^{-1}$ to obtain $A_{750-835}$ [Fig.~\ref{fig:rBN-related-Raman-feature}(b)]. This interval was chosen to encompass the broad 760--825~cm$^{-1}$ feature reported for predominantly rBN crystallites while allowing for modest spectral variation across the present Raman maps.\cite{Desrat-PhysRevMater-9-124001} A weaker Raman feature near 800~cm$^{-1}$ has also been reported in few-layer hBN and was tentatively attributed to surface-induced activation of normally infrared-active or silent out-of-plane modes.\cite{Laleyan-OE-26-23031} The corresponding $A_{2u}$ and $B_{1g}$ phonon modes of hBN in this spectral range are supported by lattice-dynamical measurements and calculations.\cite{Serrano-PRL-98-095503} However, the hBN-related feature reported near 800~cm$^{-1}$ is comparatively weak and spectrally confined, whereas the response associated with predominantly rBN material extends broadly over the 760--825~cm$^{-1}$ range. Such a localized hBN contribution is therefore expected to contribute only modestly to $A_{750-835}$ and is unlikely to account for the broad low-frequency responses observed here. Accordingly, $A_{750-835}$ is used as an absolute measure of rBN-related low-frequency Raman response, while allowing for a possible minor contribution from symmetry-activated hBN modes.

The total Raman response of the main BN band, A$_{1367}^{total}$, is obtained by integrating the experimental spectrum from $\omega_0$-100 to $\omega_0$+150~cm$^{-1}$, where $\omega_0$ is the fitted center of the $\sim$1367~cm$^{-1}$ band [Fig.~\ref{fig:rBN-related-Raman-feature}(c)]. This quantity provides a measure of the overall Raman response associated with the principal sp$^2$-BN band and is used to compare the strength of BN-related Raman scattering across the growth conditions.

The far high-wavenumber residual response, $T_{\mathrm{high}}^{\mathrm{far}}$, is obtained from the line shape of the same $\sim$1367~cm$^{-1}$ band.
 A symmetric pseudo-Voigt function fitted to the central $\pm$10~cm$^{-1}$ region defines the core response, I$_{\mathrm{core}}$, and the signed residual I$_{\mathrm{exp}}$-I$_{\mathrm{core}}$ is integrated from $\omega_0$+50 to $\omega_0$+150~cm$^{-1}$ [Fig.~\ref{fig:rBN-related-Raman-feature}(c)]. This quantity captures the extended high-wavenumber residual associated with the asymmetric tail reported for predominantly rBN material. Because the microscopic origin of this tail is less firmly established than that of the broad low-frequency response, T$_{\mathrm{high}}^{\mathrm{far}}$ is treated as a complementary rBN-related Raman observable rather than as a uniquely rBN-specific measure.

The three quantities therefore provide complementary information: $A_{1367}^{\mathrm{total}}$ tracks the overall BN Raman response, whereas $A_{750-835}$ and $T_{\mathrm{high}}^{\mathrm{far}}$ quantify two spectral features associated with rBN-related character. Their absolute magnitudes can also be influenced by crystallite orientation, texture, thickness, morphology, disorder, and optical sampling, and they are therefore not interpreted as direct measures of rBN phase fraction. Applying identical pixel-wise definitions throughout the 44-condition dataset enables the processing dependence of these Raman responses to be compared consistently.

\subsection{Temperature--composition dependence of absolute BN Raman responses}

\begin{figure}
\includegraphics[width=1.0\columnwidth]{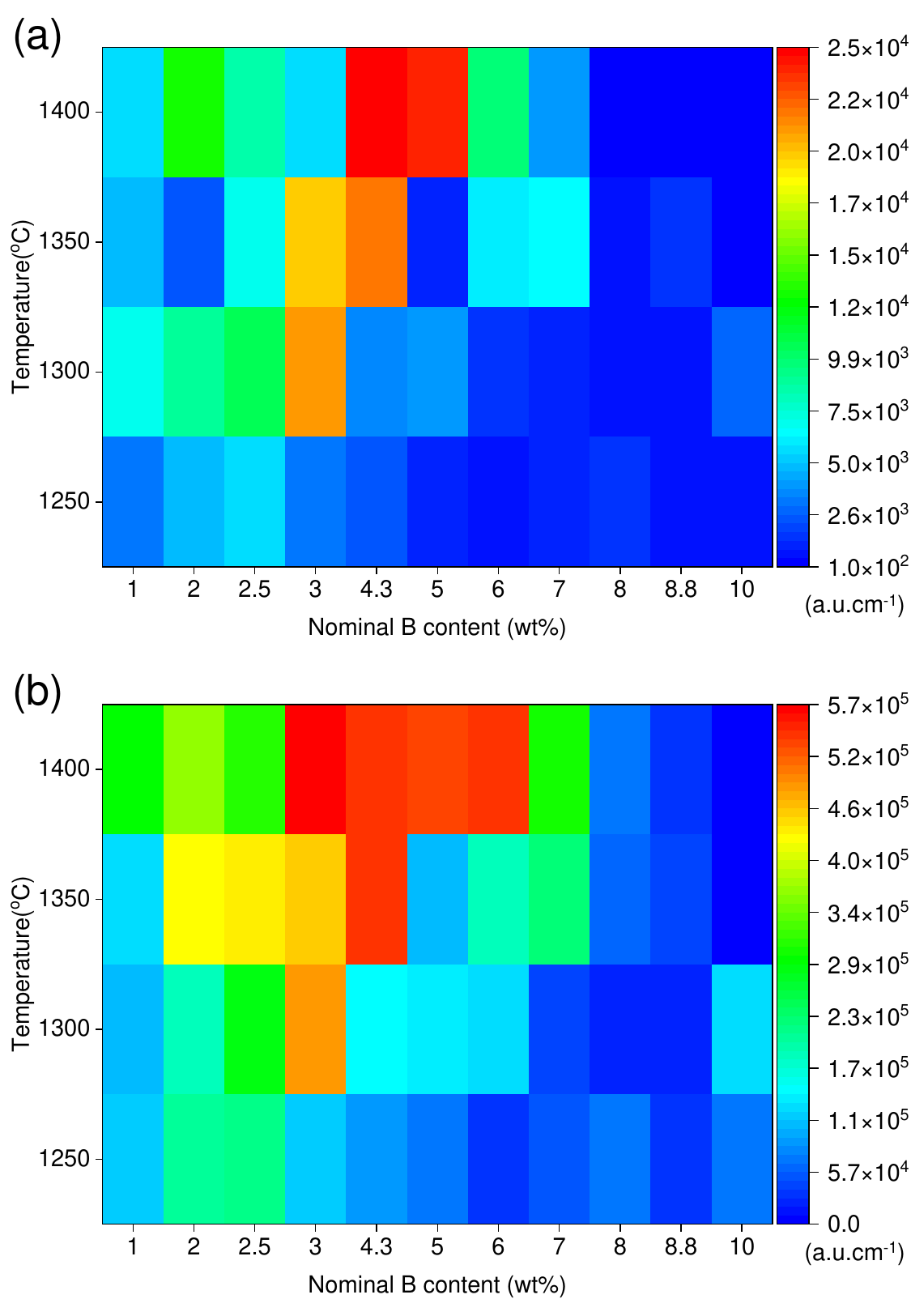}
\caption{
Temperature--composition maps of the absolute Raman responses for BN formed on Fe--B sources under flowing N$_2$. (a) Integrated low-frequency Raman response, $A_{750-835}$, and (b) total Raman response of the $\sim$1367~cm$^{-1}$ BN band, $A_{1367}^{\mathrm{total}}$, across the 44 investigated growth conditions. Each value represents the mean of the available map-level medians obtained using the nine-location Raman mapping protocol. Color scales indicate the integrated Raman areas in a.u.cm$^{-1}$.
}
\label{fig:Heatmap-LF-HF}
\end{figure}

Figure~\ref{fig:Heatmap-LF-HF} summarizes the temperature--composition dependence of the two absolute Raman responses across the 44 growth conditions. The low-frequency response, $A_{750-835}$ [Fig.~\ref{fig:Heatmap-LF-HF}(a)], is strongly enhanced within a relatively restricted region of the growth map. The strongest $A_{750-835}$ response occurs at 1400~$^\circ$C and 4.3~wt\% B, and the high-response conditions are concentrated along a relatively narrow diagonal band in the temperature--composition space, extending from approximately 1300~$^\circ$C/3~wt\% B to 1400~$^\circ$C/4.3--5~wt\% B.  The total $\sim$1367~cm$^{-1}$ BN response, $A_{1367}^{\mathrm{total}}$ [Fig.~\ref{fig:Heatmap-LF-HF}(b)], exhibits a broadly similar processing dependence. Thus, the conditions producing the strongest rBN-related low-frequency Raman response largely overlap those yielding a strong overall BN Raman response, identifying a common temperature--composition window favorable for substantial BN formation with pronounced rBN-related Raman character.

The two response maps are nevertheless not identical. Although their high-response regions overlap substantially, the locations and spatial extents of their maxima are slightly shifted relative to one another, indicating that $A_{750-835}$ does not simply scale with $A_{1367}^{\mathrm{total}}$ across the growth space. This relationship is examined directly below.

\subsection{Relationship between the low-frequency and total BN Raman responses}

\begin{figure}
\includegraphics[width=1.0\columnwidth]{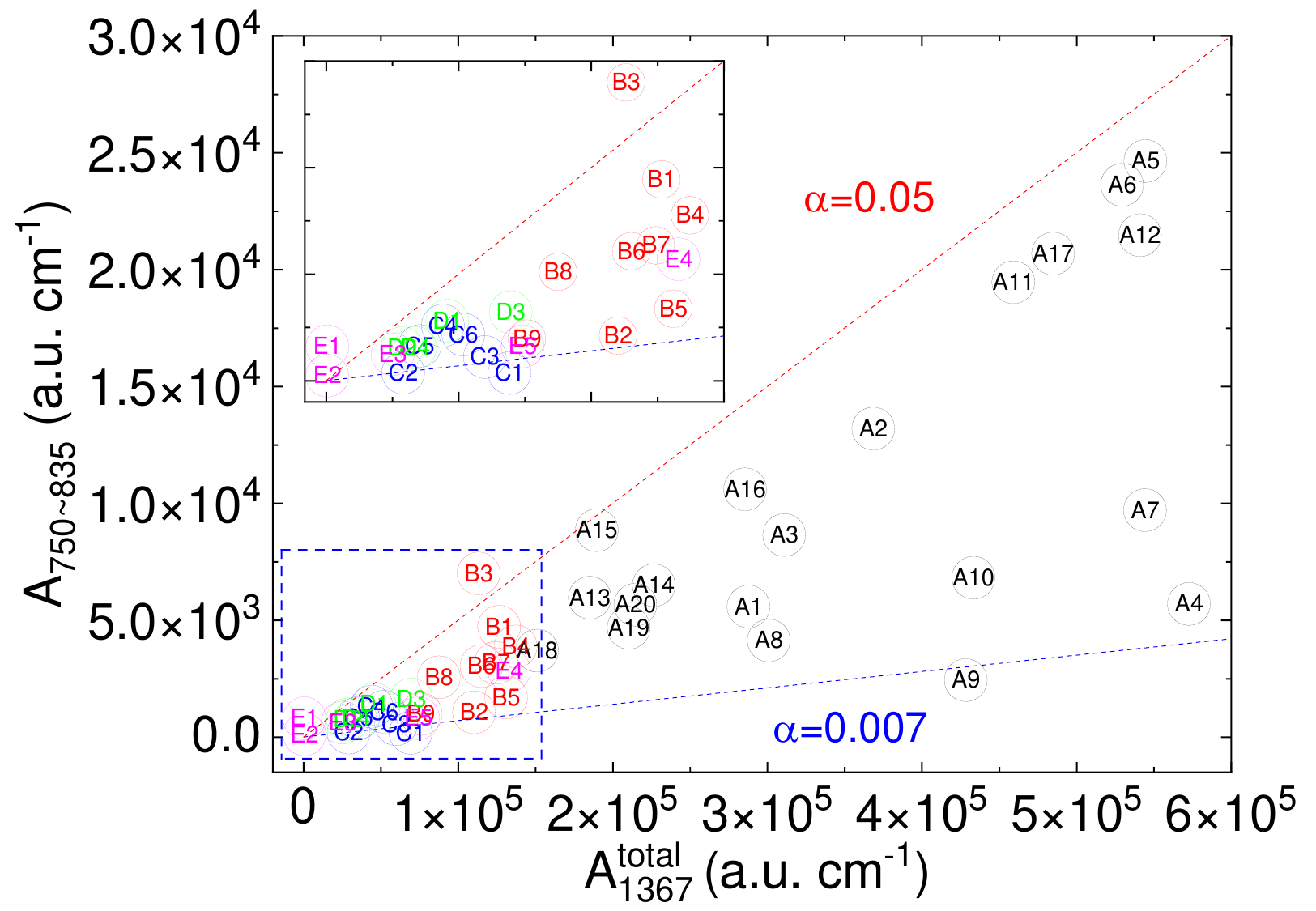}
\caption{
Relationship between the total $\sim$1367~cm$^{-1}$ BN Raman response, $A_{1367}^{\mathrm{total}}$, and the low-frequency Raman response, $A_{750-835}$, for the 44 investigated growth conditions. Each symbol represents one specimen using the mean of the available map-level median values obtained from the nine-location Raman mapping protocol. The inset enlarges the low-response region enclosed by the blue dashed box in the main panel. Red and blue dashed lines are included as guides to the relative low-frequency response and indicate constant slopes, $\alpha=A_{750-835}/A_{1367}^{\mathrm{total}}$, with $\alpha=0.05$ and 0.007, respectively.
}
\label{fig:Dependence-between-A1367-and-A750}
\end{figure}

Figure~\ref{fig:Dependence-between-A1367-and-A750} directly compares $A_{750-835}$ with $A_{1367}^{\mathrm{total}}$ for all 44 specimens. The magnitude of both Raman responses is broadly consistent with the A--E classification established from the growth-product characteristics: specimens belonging to different classes occupy characteristic regions of the plot, extending from weak responses near the origin to the much stronger responses observed predominantly for class A. The enlarged low-response region in the inset further resolves the distributions of the B--E specimens that are compressed near the origin in the main panel.

Importantly, the data do not collapse onto a single line through the origin. Instead, most specimens span relative-response slopes of approximately $\alpha=0.007$--0.05, where $\alpha=A_{750-835}/A_{1367}^{\mathrm{total}}$. These values are used only as empirical guides to the observed range of relative low-frequency response and are not interpreted as physical bounds. This provides a direct representation of the slight displacement between the high-response regions in the two temperature--composition maps in Fig.~\ref{fig:Heatmap-LF-HF}: as the growth condition changes, the two Raman responses evolve in a coordinated but not strictly proportional manner. The resulting variation in slope therefore reflects a condition-dependent change in the relative prominence of the low-frequency response, rather than only a change in the overall Raman-response magnitude.

A similar range of slopes is observed within several of the A--E classes rather than being associated with only one particular group. Thus, the A--E classification is reflected primarily in the absolute magnitudes of the Raman responses, whereas variation in their relative contributions persists across much of the growth map.

\subsection{Strong correspondence between the low-frequency Raman response and the far high-wavenumber residual response}

\begin{figure}
\includegraphics[width=1.0\columnwidth]{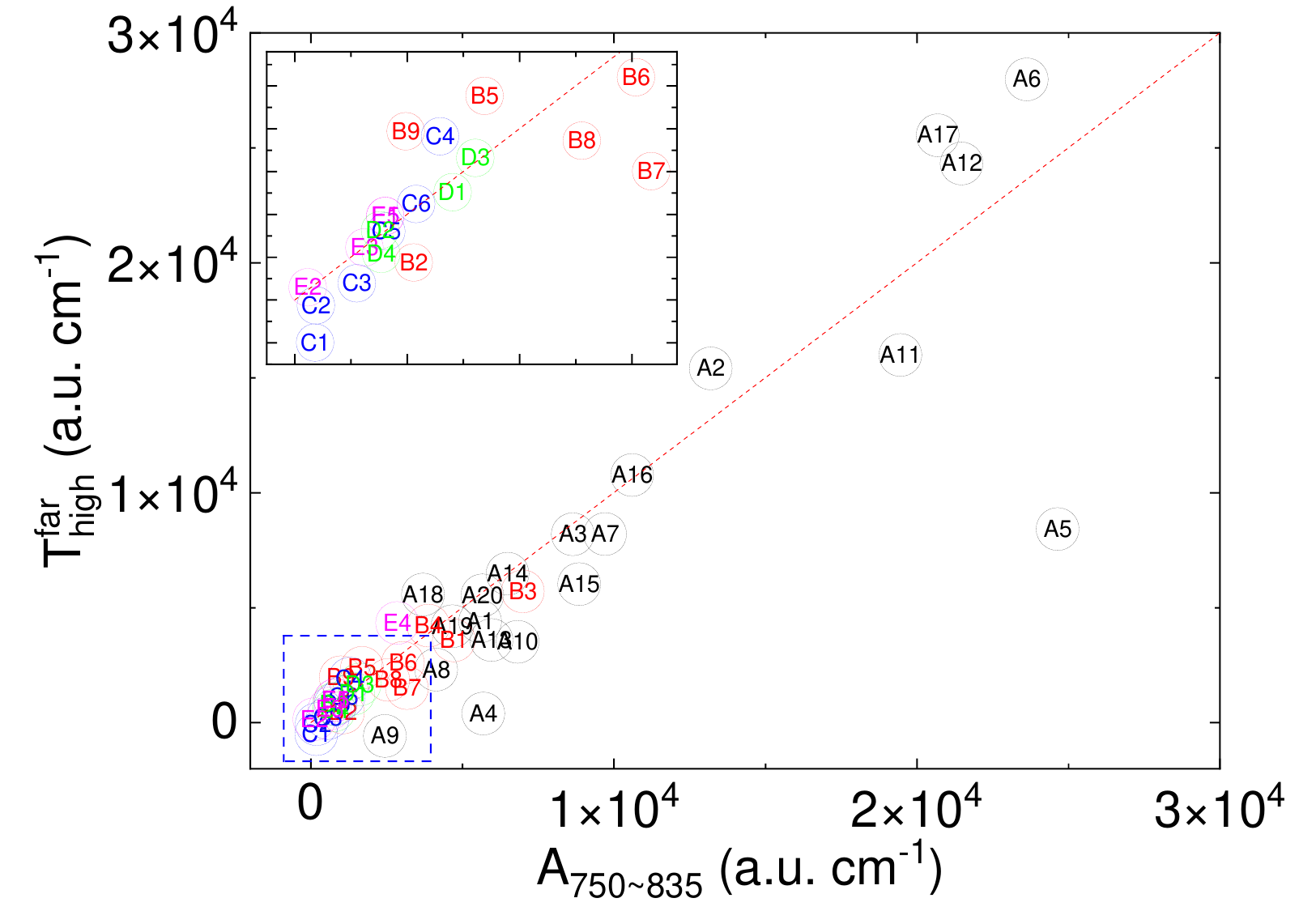}
\caption{
Relationship between the low-frequency Raman response, $A_{750-835}$, and the far high-wavenumber residual response, $T_{\mathrm{high}}^{\mathrm{far}}$, for the 44 investigated growth conditions. Each symbol represents one specimen using the mean of the available map-level median values obtained from the nine-location Raman mapping protocol. The red dashed line is a one-to-one reference line, $T_{\mathrm{high}}^{\mathrm{far}}=A_{750-835}$. The inset enlarges the low-response region enclosed by the blue dashed box in the main panel.
}
\label{fig:Dependence-between-A750-and-T_far-high}
\end{figure}

Figure~\ref{fig:Dependence-between-A750-and-T_far-high} compares the low-frequency Raman response, $A_{750-835}$, with the far high-wavenumber residual response, $T_{\mathrm{high}}^{\mathrm{far}}$, across all 44 specimens. In contrast to the broader spread observed between $A_{750-835}$ and $A_{1367}^{\mathrm{total}}$, these two responses exhibit a strong correspondence over nearly the entire measured range, with most specimens lying near the one-to-one reference line shown in Fig.~\ref{fig:Dependence-between-A750-and-T_far-high}. To quantify this relationship, a linear regression constrained through the origin was applied, yielding $T_{\mathrm{high}}^{\mathrm{far}}=0.907\times A_{750-835}$ with $R^2=0.81$ and Pearson $r=0.90$. Here, $R^2$ was evaluated using the conventional mean-centered total sum of squares and represents the fraction of the variance in $T_{\mathrm{high}}^{\mathrm{far}}$ described by the constrained linear model, whereas Pearson $r$ quantifies the strength and direction of the linear association from the normalized co-variation of the two responses about their respective means. These values quantitatively support the strongly coordinated evolution of $A_{750-835}$ and $T_{\mathrm{high}}^{\mathrm{far}}$ across the 44 growth conditions and the A--E specimen classes.

This strong correspondence is consistent with the two rBN-related features tracking closely related spectral evolution rather than simply the overall BN Raman response. The two quantities are nevertheless not strictly equivalent, as indicated by finite deviations from the one-to-one reference line and by the dependence of the absolute $T_{\mathrm{high}}^{\mathrm{far}}$ magnitude on the selected integration window. This relationship is also robust to reasonable variations in the integration limits: varying the lower and upper offsets over $+40$ to $+60$ and $+125$ to $+175$~cm$^{-1}$, respectively, gives Pearson $r=0.88$--0.92 and mean-centered $R^2=0.78$--0.84, while the resulting specimen-level $T_{\mathrm{high}}^{\mathrm{far}}$ values remain highly correlated with those obtained using the reference $+50$ to $+150$~cm$^{-1}$ window ($r\geq0.996$).

\subsection{Growth-history dependence and Fe--B composition context of rBN-related Raman character}

The temperature--composition series shows that the low-frequency Raman response does not scale with the total BN Raman response through a single proportionality, but instead exhibits a condition-dependent relative contribution. The dwell-time experiment extends this observation into the time domain by varying the duration of the high-temperature N$_2$ exposure while keeping the growth temperature, nominal B content, and subsequent cooling protocol unchanged. Fe--B sources containing 8, 8.8, and 10~wt\% B were therefore held at 1350~$^\circ$C under N$_2$ for either 60 or 300~min. The corresponding 60-min specimens are C3, D1, and E2, respectively.

\begin{figure}
\includegraphics[width=1.0\columnwidth]{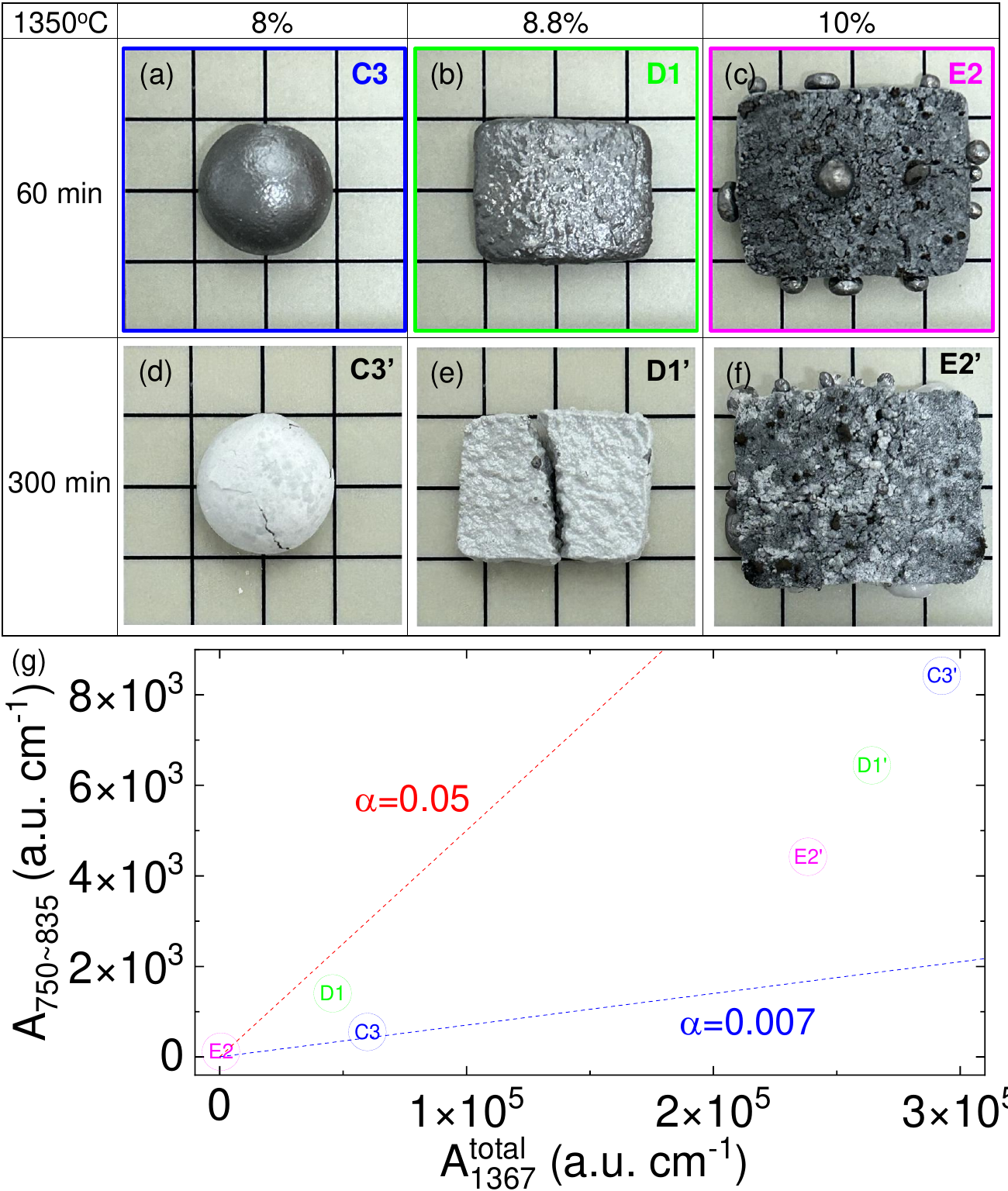}
\caption{Growth-duration dependence of the macroscopic surface appearance and Raman responses of Fe--B sources at 1350~$^\circ$C. Representative photographs of specimens with nominal B contents of 8, 8.8, and 10~wt\% after growth under flowing N$_2$ for (a--c) 60~min (C3, D1, and E2) and (d--f) 300~min (C3$'$, D1$'$, and E2$'$), respectively. The 60-min specimens correspond to the samples identified in Fig.~\ref{fig:sample-images}. Each photograph includes a background grid for dimensional reference, with four grid squares corresponding to 26~mm in both the horizontal and vertical directions. (g) Relationship between the total $\sim$1367~cm$^{-1}$ BN Raman response, $A_{1367}^{\mathrm{total}}$, and the low-frequency Raman response, $A_{750-835}$, for the corresponding 60- and 300-min specimens. Each plotted value represents the arithmetic mean of the available map-level median responses obtained using the nine-location Raman mapping protocol. Dashed lines indicate constant slopes, $\alpha=A_{750-835}/A_{1367}^{\mathrm{total}}$, of 0.007 and 0.05 and are included only as empirical guides to the relative prominence of the low-frequency response.}
\label{fig:Growth-duration-dependence}
\end{figure}

The macroscopic appearance of the sources changes substantially during the prolonged dwell [Fig.~\ref{fig:Growth-duration-dependence}(a--f)]. At 8 and 8.8~wt\% B, extending the dwell from 60 to 300~min produces a pronounced increase in apparent film-like surface coverage, whereas the 10~wt\% B source retains a more heterogeneous surface appearance. These observations show that the Fe--B source continues to evolve during the high-temperature N$_2$ stage and that the extent of this evolution depends on nominal B content.

The corresponding Raman responses also change markedly with dwell time [Fig.~\ref{fig:Growth-duration-dependence}(g)]. For the 8 and 8.8~wt\% B specimens, both $A_{750-835}$ and $A_{1367}^{\mathrm{total}}$ increase strongly after 300~min, producing large displacements toward higher response magnitudes. The 10~wt\% B specimen likewise changes, although the magnitude of the shift is smaller. Importantly, the 60- and 300-min specimens do not follow a single common slope in the $A_{750-835}$--$A_{1367}^{\mathrm{total}}$ representation. Thus, prolonging the dwell changes not only the overall Raman-response magnitude but also the relative contribution of the low-frequency response, demonstrating that growth duration constitutes an additional kinetic variable whose effect depends on source composition.

These results can be placed in the context of previous BN-growth studies using Fe--B sources. Desrat \textit{et al.} reported predominantly rBN crystallites from an Fe--2.5~wt\% B flux treated at 1550--1600~$^\circ$C under N$_2$ followed by very slow cooling.\cite{Desrat-PhysRevMater-9-124001} In contrast, Shi \textit{et al.} obtained multilayer hBN from Fe$_2$B (approximately 8.8~wt\% B) at 1100--1300~$^\circ$C,\cite{Shi-RSCAdv-9-10155} and from a molten Fe$_{82}$B$_{18}$ alloy (approximately 4.1~wt\% B) at 1250~$^\circ$C under isothermal growth conditions.\cite{Shi-NC-11-849} These studies show that different BN stacking outcomes can be obtained from Fe--B sources spanning substantially different compositions and thermal histories, without establishing a particular B concentration as intrinsically optimal for rBN formation.

The present temperature--composition mapping extends this comparison by systematically resolving the rBN-related Raman response over a broad composition range. The strongest $A_{750-835}$ responses are not associated with a single nominal B content, but shift from approximately 3~wt\% B at 1300~$^\circ$C toward 4.3--5~wt\% B at 1400~$^\circ$C. The 3--5~wt\% B range corresponds to approximately 13.8--21.4~at\% B and partly overlaps the Fe-rich eutectic composition near 3.8~wt\% B ($\sim$17~at\% B).\cite{Liao-PD-41-1993,Okamoto-JPED-25-297,Rompaey-JAC-334-173} This overlap provides useful thermodynamic context for the composition dependence, but does not establish the eutectic composition itself as the origin of the enhanced rBN-related response. In particular, the shift of the high-response region toward higher B content with increasing temperature indicates that a fixed B concentration alone cannot account for the observed behavior.

Taken together, the temperature--composition mapping and dwell-time results show that the final rBN-related Raman character cannot be described by nominal B content or growth temperature alone. The dependence on dwell time further demonstrates that the Fe--B/BN system continues to evolve during high-temperature N$_2$ exposure and that this evolution is composition dependent. The contrasting rBN and hBN outcomes previously reported for different Fe--B compositions and thermal protocols are consistent with such a coupled dependence on composition and thermal history.\cite{Desrat-PhysRevMater-9-124001,Shi-RSCAdv-9-10155,Shi-NC-11-849} Although the present measurements do not identify the specific microscopic variable controlling BN stacking, they establish that conditions favorable for strong rBN-related Raman response form a restricted temperature–composition window, with the Raman response additionally modulated by growth duration/history, rather than a universal optimum B concentration.

\subsection{Spatial heterogeneity and robustness of specimen-level Raman responses}

The macroscopic surface appearance in Fig.~\ref{fig:sample-images} indicates substantial spatial heterogeneity within individual Fe--B sources, making the choice of a specimen-level Raman response nontrivial. The nine-location Raman mapping protocol was therefore used not only to sample this heterogeneity but also to evaluate whether the resulting spatial distributions could be represented reproducibly by a single specimen-level value. For a Raman quantity $Q$, let $M_{s,j}^{Q}$ denote the median value obtained from the analysis-valid pixels in map $j$ of specimen $s$. The specimen-level value used throughout this work was calculated as the arithmetic mean of the available map-level medians,

$$
\overline{M}_{s}^{Q}=\frac{1}{n_s}\sum_{j=1}^{n_s}M_{s,j}^{Q},
$$

where $n_s$ is the number of maps providing an analysis-valid value, normally nine. This hierarchical procedure first reduces the influence of isolated pixel-level extremes through the map median and then assigns equal weight to the spatially separated sampling locations.

For the positive absolute-response quantities, $Q=A_{750-835}$ or $A_{1367}^{\mathrm{total}}$, the degree of spatial variation within each specimen was quantified using the map-to-map coefficient of variation,

$$
\mathrm{CV}_{s}^{Q}=
\frac{\sigma_{s}^{Q}}{\overline{M}_{s}^{Q}},
\qquad
\sigma_{s}^{Q}=
\sqrt{
\frac{1}{n_s-1}
\sum_{j=1}^{n_s}
\left(M_{s,j}^{Q}-\overline{M}_{s}^{Q}\right)^2
},
$$

and the overall magnitude of spatial heterogeneity was summarized by the median of $\mathrm{CV}_{s}^{Q}$ across the 44 baseline growth conditions. Considerable map-to-map variation was observed for the absolute Raman responses: the median map-to-map coefficients of variation were approximately 0.47 for $A_{750-835}$ and 0.41 for $A_{1367}^{\mathrm{total}}$. Thus, substantial spatial heterogeneity remains even after pixel-level fluctuations within each map are represented by their medians.

Despite this heterogeneity, the condition-to-condition trends in the specimen-level responses were robust to reasonable choices of the method used to aggregate the spatially distributed map-level measurements. As an alternative to the mean of the map-level medians, $\overline{M}_{s}^{Q}$, we also calculated their median,

$$
\widetilde{M}_{s}^{Q}
=
\operatorname{median}_{j}
\left(M_{s,j}^{Q}\right).
$$

The values obtained using $\overline{M}_{s}^{Q}$ were strongly correlated with $\widetilde{M}_{s}^{Q}$, with Pearson $r=0.996$, 0.994, and 0.993 for $A_{750-835}$, $A_{1367}^{\mathrm{total}}$, and $T_{\mathrm{high}}^{\mathrm{far}}$, respectively.

Moreover, the relationship between $A_{750-835}$ and $T_{\mathrm{high}}^{\mathrm{far}}$ was essentially unchanged when the specimen-level aggregation method was varied: the origin-constrained fits gave slopes of 0.907--0.940, $R^2=0.81$--0.82, and Pearson $r=0.90$--0.91. Importantly, directly pairing the map-level medians $M_{s,j}^{A_{750-835}}$ and $M_{s,j}^{T_{\mathrm{high}}^{\mathrm{far}}}$ from the same spatial locations also produced a strong correspondence ($r=0.89$), demonstrating that the specimen-level relationship does not arise from independently averaging spatially heterogeneous signals. These results show that, although the Fe--B source surfaces are spatially heterogeneous, the hierarchical map-median/specimen-mean procedure provides a robust specimen-level representation of the Raman responses associated with each growth condition.

\subsection{Complementary structural evidence from powder XRD and SHG}

\begin{figure}
\includegraphics[width=1.0\columnwidth]{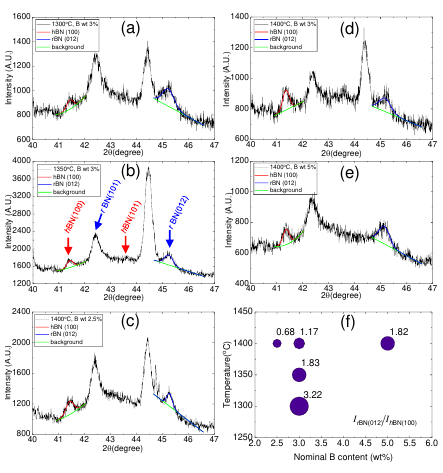}
\caption{
Powder XRD analysis of selected BN-containing products formed on Fe--B sources. The 40--47$^\circ$ regions are shown for specimens grown at (a) 1300~$^\circ$C with 3~wt\% B, (b) 1350~$^\circ$C with 3~wt\% B, (c) 1400~$^\circ$C with 2.5~wt\% B, (d) 1400~$^\circ$C with 3~wt\% B, and (e) 1400~$^\circ$C with 5~wt\% B. The black curves show the measured diffraction patterns, the red and blue curves show the fitted hBN(100) and rBN(012) profiles, respectively, and the green curves show the corresponding local backgrounds. The arrows in (b) indicate the expected Cu K$\alpha$ diffraction positions of hBN(100), rBN(101), hBN(101), and rBN(012) near 41.6$^\circ$, 42.6$^\circ$, 43.9$^\circ$, and 45.6$^\circ$, respectively. The prominent diffraction feature near 44.5$^\circ$ is attributed primarily to residual $\alpha$-Fe(110) from the Fe--B source, with a possible contribution from Fe$_2$B, and is not included in the BN peak analysis. (f) Temperature--composition representation of the integrated-intensity ratio $Q_{\mathrm{XRD}}=I_{\mathrm{rBN}(012)}/I_{\mathrm{hBN}(100)}$ for the five XRD-quantifiable specimens, with the area of each circle proportional to $Q_{\mathrm{XRD}}$ and the corresponding numerical value indicated adjacent to each symbol.
}
\label{fig:powder-XRD}
\end{figure}

Powder XRD and polarization-dependent SHG were used as complementary structural probes, providing stacking-sensitive and symmetry-sensitive support, respectively, for the Raman-based interpretation. Powder XRD was performed on a subset of the 44 growth conditions from which sufficient BN-containing material could be mechanically recovered. The selection was based on practical powder recoverability rather than Raman-response magnitude. Figure~\ref{fig:powder-XRD}(a--e) shows powder XRD patterns in the 40--47$^\circ$ region for specimens grown at 1300~$^\circ$C/3~wt\% B, 1350~$^\circ$C/3~wt\% B, 1400~$^\circ$C/2.5~wt\% B, 1400~$^\circ$C/3~wt\% B, and 1400~$^\circ$C/5~wt\% B. The rBN-related reflection near 42.6$^\circ$ was not used because of substantial overlap with Fe--B-related diffraction, and the comparison therefore focuses on the resolvable hBN(100) and rBN(012) reflections. The intense feature near 44.5$^\circ$, originating predominantly from residual Fe-containing source material, was excluded from the BN-specific analysis.

All five specimens exhibit a resolvable rBN(012) contribution, providing independent diffraction evidence for rBN-related stacking within the BN products and supporting the assignment of the broad 750--835~cm$^{-1}$ Raman response as rBN-related. For semi-quantitative comparison, the relative diffraction intensity $Q_{\mathrm{XRD}}=I_{\mathrm{rBN}(012)}/I_{\mathrm{hBN}(100)}$ was evaluated for the five specimens. The resulting values vary substantially with growth condition but do not show a simple ordering with the relative prominence of the low-frequency Raman response. Such one-to-one correspondence is not expected because Raman mapping probes spatially resolved regions of the intact surface and is sensitive to local optical response, orientation, morphology, and stacking-related spectral character, whereas powder XRD averages mechanically recovered material and is additionally influenced by texture, crystallite size, stacking disorder, and the amount of material contributing to each reflection. The diffraction ratio and Raman descriptors should therefore not be interpreted as interchangeable measures of rBN content.

\begin{figure}
\includegraphics[width=1.0\columnwidth]{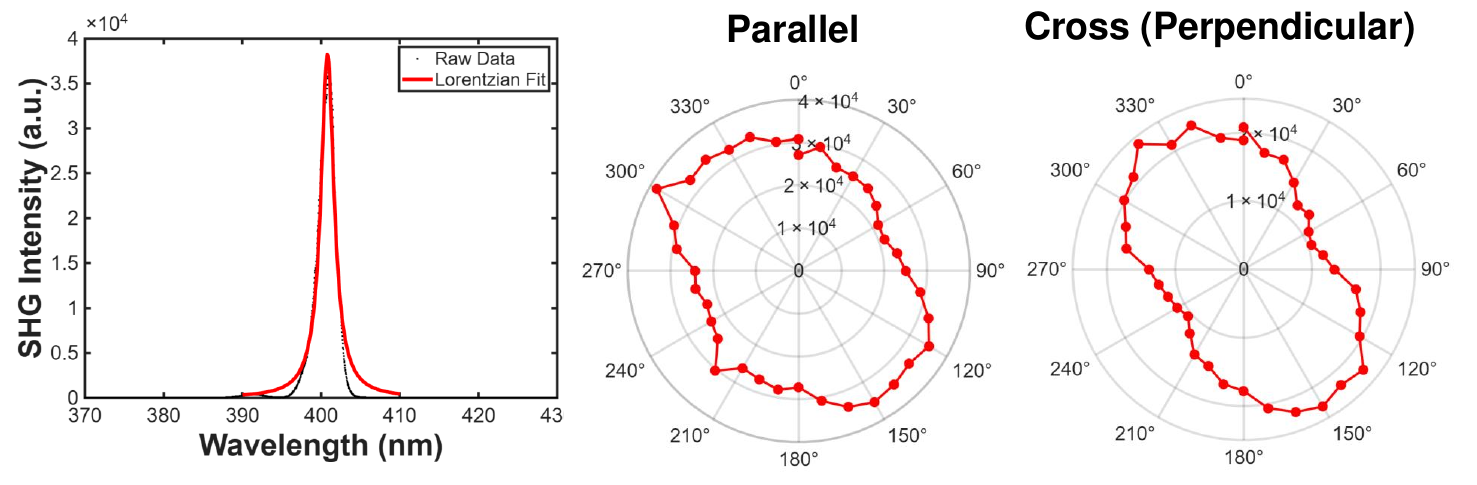}
\caption{Polarization-dependent second-harmonic generation (SHG) response of a wet-etch-separated, relatively thick BN product grown at 1350~$^\circ$C with 4.3~wt\% B. A representative SHG spectrum shows a clear emission peak near 400~nm together with a Lorentzian fit. The angular dependences of the SHG intensity are shown for parallel and cross (perpendicular) polarization configurations. The measurement provides complementary symmetry-sensitive evidence for the BN product.}
\label{fig:SHG}
\end{figure}

Figure~\ref{fig:SHG} shows the polarization-dependent SHG response of a wet-etch-separated, relatively thick BN product grown at 1350~$^\circ$C with 4.3~wt\% B, a condition included in the temperature--composition dataset. A clear emission peak near 400~nm is observed, and the SHG intensity varies systematically with polarization angle in both the parallel and cross-polarized configurations. Because rBN lacks inversion symmetry whereas ideal bulk AA$'$-stacked hBN is centrosymmetric, the observed response is consistent with an rBN-related or otherwise locally symmetry-broken BN contribution. Unlike the clear six-fold SHG symmetry reported for well-oriented single-crystalline rBN,\cite{Wang-Nature-629-74} however, the present angular response does not exhibit a well-defined six-fold pattern. This difference may reflect orientational averaging among multiple domains or spatial heterogeneity within the relatively thick separated product; accordingly, the angular dependence is not used here to assign a specific crystallographic orientation or domain structure. SHG intensity and angular dependence can also depend on active thickness or volume, coverage, optical interference, morphology, and collection efficiency, and are therefore not interpreted quantitatively in terms of rBN phase fraction or growth yield.

Taken together, XRD and SHG provide complementary structural support for the Raman-based identification of rBN-related character. The resolvable rBN(012) reflections directly establish the presence of rBN-related stacking in the selected powder products, whereas SHG independently demonstrates a non-centrosymmetric optical contribution in the 1350~$^\circ$C/4.3~wt\% B product. Because the three measurements probe different structural attributes and averaging scales, quantitative correspondence among their signal magnitudes is neither assumed nor required. Together, the three measurements provide spectral, diffraction, and symmetry-sensitive evidence for rBN-related BN structure.

\section{Conclusions}

This work establishes a temperature--composition landscape for rBN-related Raman character in BN formed on Fe--B sources under flowing N$_2$. Strong low-frequency rBN-related response occurs within a restricted growth window that substantially overlaps the region of enhanced overall BN Raman response, although the two do not follow a single proportionality. A central contribution is the independent quantification of the far high-wavenumber residual response, $T_{\mathrm{high}}^{\mathrm{far}}$, which shows strong correspondence with the low-frequency response $A_{750-835}$ across the full dataset. Their coordinated evolution supports their use as complementary Raman observables of rBN-related structural character. Powder XRD and polarization-dependent SHG provide independent stacking- and symmetry-sensitive support for this interpretation.

The growth-duration experiments further show that the Raman response depends on high-temperature dwell time as well as temperature and composition. Together, these results indicate that strong rBN-related character is associated with a restricted temperature--composition window rather than with a universal B composition, while its magnitude and relative spectral character are further modulated by growth duration. This behavior points to the evolving Fe--B/BN system during high-temperature N$_2$ exposure as an important factor influencing the final BN structural response. Resolving the microscopic origin of this evolution will require correlative local structural measurements and in situ or source-state-sensitive characterization.

\section*{Acknowledgments}
This work was supported by the National Research Foundation of Korea (NRF) grant funded by the Korea government (MSIT) (RS-2021-NR060087, RS-2023-00240724, RS-2025-25457100, RS-2022-NR068228, RS-2026-25509069, IITP-2026-RS-2024-00437191) and through Korea Basic Science Institute (National Research Facilities and Equipment Center) grant (2021R1A6C101A437) funded by the Ministry of Education.



%

\end{document}